\documentclass[12pt]{article}

\usepackage[utf8]{inputenc}
\usepackage[T1]{fontenc}
\usepackage{graphicx}
\usepackage{adjustbox}
\usepackage[authoryear,round]{natbib}
\usepackage{algorithm2e}
\usepackage{algpseudocode}
\usepackage{xcolor}
\usepackage{listings}
\usepackage{booktabs}
\usepackage{multirow}
\usepackage{makecell}
\usepackage[colorlinks=true,linkcolor=blue,citecolor=blue,urlcolor=blue]{hyperref}
\usepackage{colortbl}
\usepackage{amsmath,amssymb,amsfonts}
\usepackage{amsthm}
\usepackage{mathrsfs}
\usepackage[title]{appendix}
\usepackage{textcomp}
\usepackage{algorithmicx}
\usepackage{authblk}
\usepackage{float}
\usepackage[section]{placeins}
\usepackage{geometry}

\title{LLM-based vs. Search-based Merge Conflict Resolution:\\
An Empirical Study of Competing Paradigms}

\author[1]{Heleno de Souza Campos Junior}
\author[1]{Leonardo Gresta Paulino Murta}

\affil[1]{Department of Computer Science, Federal Fluminense University,\\
Av. Gal. Milton Tavares de Souza, Niterói, RJ 24210-346, Brazil\\
\texttt{helenocampos@id.uff.br, leomurta@ic.uff.br}}

\date{}

\begin{document}

\maketitle

\vspace{1em}

\begin{abstract}

\textbf{Context:} The resolution of software merge conflicts is being reshaped by two competing and novel paradigms: generative approaches based on Large Language Models (LLMs) and optimization approaches from Search-Based Software Engineering (SBSE). While tools from both paradigms have shown promise, their relative strengths, weaknesses, and fundamental trade-offs are not yet understood.

\textbf{Objective:} This paper presents the first in-depth empirical study to directly compare these two paradigms, aiming to identify their respective capabilities and limitations in real-world scenarios.

\textbf{Method:} We evaluated MergeGen, a state-of-the-art LLM-based tool, against SBCR --- a novel SBSE approach employing a Random Restart Hill Climbing (RRHC) algorithm. The comparison used thousands of real-world conflicts from open-source projects written in Java, C\#, JavaScript, and TypeScript.

\textbf{Results:} Our findings reveal fundamental trade-offs. The LLM paradigm excels at resolving conflicts with imbalanced content by leveraging learned patterns. However, it struggles with non-English content or large inputs, which can lead to truncated or empty resolutions. Conversely, the SBSE paradigm demonstrates superior generalization across datasets and performs optimally on balanced conflicts, showcasing its potential as a robust, data-independent alternative.

\textbf{Conclusions:} Neither paradigm is a silver bullet. Our findings highlight context-dependent strengths and advocate for the development of hybrid systems that intelligently leverage the complementary capabilities of both LLM and SBSE approaches. This will help create more robust and reliable merge conflict resolution tools.

\end{abstract}


\newpage

\section{Introduction}

In modern collaborative software development, version control systems play a crucial role in managing changes across multiple developers working concurrently. These systems allow developers to work independently, merging their contributions into a shared codebase. However, the merging of code changes often leads to conflicts when concurrent modifications are made to overlapping code regions. Such merge conflicts must be manually resolved, which can be both time-consuming and error-prone, increasing the risk of introducing defects \citep{grinter1996supporting, shihab2012effect}. Automated merge conflict resolution techniques aim to alleviate this burden by generating merge resolutions that align closely with developers' expectations, reducing manual effort and the potential for errors \citep{mens2002state, cavalcantiEvaluatingImprovingSemistructured2017o}.

A crucial insight from large-scale empirical studies is that the vast majority of merge conflicts do not require developers to write new code. \citet{ghiotto2020nature} found that in 87\% of cases, the resolution is composed entirely of lines already present in the conflicting versions. These resolutions can be categorized into two groups: trivial and non-trivial. Trivial resolutions follow simple patterns, such as selecting one of the conflicting versions in its entirety or concatenating them. The remaining cases require a non-trivial combination of lines from both versions, which can be seen as a cherry-picking process that interleaves content, eventually adding new code. While writing new code (13\% of cases) is also a non-trivial strategy, the combination strategy (9\% of cases) is of special interest because of its combinatorial nature. Finding the correct sequence of lines among all possible permutations is the most difficult resolution task that can be solved without adding new code.

To tackle this long-standing challenge, the research community has explored various techniques, from structured merging to machine learning. Recently, this has culminated in the emergence of two competing and novel paradigms. The first is a generative approach based on Large Language Models (LLMs), represented by tools like MergeGen \citep{dong2023merge}, which leverage vast training data to produce resolutions. The second is an optimization-based approach from Search-Based Software Engineering (SBSE), represented by the SBCR approach \citep{camposjunior2025sbcr}, which frames conflict resolution as a combinatorial search problem solved by heuristic algorithms.

While tools from both paradigms have shown promise, their relative strengths, weaknesses, and fundamental trade-offs are not yet understood. The emergence of these two distinct and modern paradigms---one data-driven and generative, the other heuristic-driven and combinatorial---motivates a critical question: what are their comparative capabilities and ideal use cases?

This paper presents the first in-depth empirical study to directly compare these two paradigms. We evaluate MergeGen \citep{dong2023merge}, a state-of-the-art LLM-based tool, against SBCR \citep{camposjunior2025sbcr}, which employs a Random Restart Hill Climbing (RRHC) algorithm. Our evaluation focuses specifically on the challenging combination strategy. Scoping our study to these conflicts is deliberate: it allows us to compare the core mechanisms of both paradigms on a complex, non-trivial task that is solvable without generating new content, thus providing a fair and direct basis for comparison. Our study is guided by the following research questions:
\begin{itemize}
    \item RQ1: How does the performance of an LLM-based approach compare to a search-based approach for resolving merge conflicts?

    \item RQ2: Does the performance difference between the LLM and search-based paradigms vary across different programming languages?

    \item RQ3: How do the generalization capabilities of the two paradigms differ when evaluated on previously unseen datasets?

    \item RQ4: What explains the performance differences between the LLM and search-based paradigms?
\end{itemize}

Our experimental results reveal fundamental trade-offs. The LLM paradigm excels at resolving conflicts with imbalanced content between versions, likely benefiting from learned contextual patterns. However, it shows signs of overfitting and struggles with non-English content or large inputs, which can lead to truncated or empty resolutions. Conversely, the SBSE paradigm demonstrates superior generalization across datasets and performs optimally on balanced conflicts, showcasing its potential as a robust, data-independent alternative.

We conclude that neither paradigm offers a one-size-fits-all solution. Our findings highlight context-dependent strengths and advocate for the development of hybrid systems that intelligently leverage the complementary capabilities of both LLM and SBSE approaches. This will help create more robust and reliable conflict resolution tools.

The remainder of this paper is structured as follows: Section 2 details MergeGen and SBCR. Section 3 outlines the materials and methods used in our comparative experiments. Section 4 presents the experimental results addressing our research questions. Section 5 discusses the key findings, limitations, and implications of our study. Section 6 presents the threats to the validity of our results and our actions to mitigate them. Section 7 covers related work, and Section 8 concludes the paper.

\section{Compared Paradigms and Approaches}
\label{sec:compared_approaches}

This section details the two competing paradigms for merge conflict resolution evaluated in this study: the Generative AI (GenAI) paradigm, represented by the state-of-the-art tool MergeGen \citep{dong2023merge}, and the Search-Based Software Engineering (SBSE) paradigm, represented by the novel SBCR approach \citep{camposjunior2025sbcr}. For each, we first explain the intuition behind the paradigm's problem-solving approach before detailing the concrete implementation employed in our study.

\subsection{The Generative AI Paradigm represented by MergeGen}

The core intuition of the GenAI paradigm is to treat conflict resolution as a sophisticated sequence prediction task. An LLM approaches this problem not by analyzing combinatorial possibilities, but by leveraging patterns learned from observing millions of code changes and resolutions. It views the conflicting versions as a source sequence and aims to generate the most probable target sequence (the resolution) that logically follows from the context. This process is akin to language translation, where the model translates the conflicting inputs into a single, coherent output, with the ability to infer missing information or generate entirely new content if its training data suggests such a pattern is appropriate.

MergeGen \citep{dong2023merge} is a state-of-the-art implementation of this paradigm. It is built on two key components that enable this generative process:

\textbf{1. Structured Conflict Representation}: Instead of using raw text, MergeGen first processes the conflict at a token level. It then creates a structured input sequence using special tokens to explicitly delineate the conflicting regions and align the two conflicting versions ($V_1$ and $V_2$) with their common ancestor ($O$). This fine-grained, conflict-aware representation provides the model with a clear context of what has changed, simplifying the resolution task.

\textbf{2. Generative Model}: The second component is an encoder-decoder Transformer, specifically the CodeT5 model, which is pre-trained on a large corpus of code and fine-tuned on language-specific conflict resolution data. This model generates the final resolution token-by-token in an auto-regressive manner. This mechanism gives MergeGen the flexibility to produce novel code sequences that may not have been present in any of the input versions, making it theoretically capable of handling both \textit{Combination} and \textit{New Code} resolution strategies.

\subsection{The Search-Based Paradigm represented by SBCR}

The intuition of the SBSE paradigm is to frame conflict resolution as a search for the best solution within a vast landscape of possibilities. This approach does not attempt to understand the code's semantics. Instead, it defines a solution space composed of all possible combinations of lines from the conflicting versions. The goal is to navigate this landscape using heuristic-guided search, iteratively moving towards solutions with a better quality score, until an optimal or near-optimal resolution is found. The intelligence of this paradigm lies not in learned patterns, but in the design of a lightweight evaluation function that effectively guides the search.

SBCR \citep{camposjunior2025sbcr} is a novel implementation of this paradigm, requiring three key components:

\textbf{1. Problem Representation}: A candidate solution is represented as an ordered sequence of lines from the two conflicting versions, $V_1$ and $V_2$. The search is constrained to respect the partial order of lines within each version, a heuristic that drastically reduces the search space while covering 98.6\% of real-world combination-based resolutions \citep{camposjunior2024composition}.

\textbf{2. Evaluation Function}: A lightweight proxy for resolution quality is used, avoiding expensive compilation or testing. The function assesses a candidate's quality by calculating the \textit{mean} of its textual similarity to both parent versions ($V_1$ and $V_2$), based on the finding that this metric is moderately correlated ($\rho=0.64$) with the final resolution's quality \citep{camposjunior2025sbcr}.

\textbf{3. Manipulation Operators and Search Algorithm}: SBCR uses three fundamental line operators (Addition, Removal, Position Exchange) to explore the neighborhood of a solution. It employs a Random Restart Hill Climbing (RRHC) algorithm to navigate the search space, performing local searches to find optima and restarting from random points to avoid getting stuck.

\subsection{Summary of Paradigmatic Differences}

The descriptions of MergeGen and SBCR highlight fundamental differences in their underlying paradigms. MergeGen, representing the GenAI paradigm, is a data-driven system that learns from historical data. Its core mechanism is generative, allowing it to produce resolutions that can fall under both the \textit{Combination} and \textit{New Code} categories. It relies on models like CodeT5 and fine-tuning on language-specific datasets of conflict resolutions.

In contrast, SBCR, representing the SBSE paradigm, is a heuristic-driven optimization approach. It requires no training data; its search is guided entirely by a problem-specific evaluation function. Its mechanism is purely combinatorial, creating resolutions exclusively by selecting and ordering existing lines. This inherently limits its capability to the \textit{Combination} strategy and makes it incapable of generating \textit{New Code}. However, this also makes SBCR language-agnostic, as its logic is independent of code syntax. These core distinctions form the basis of our comparative study.

\section{Materials and methods}
\label{sec:materials}

This section details the experimental design used for the empirical comparison of the GenAI and SBSE paradigms. We first restate our research questions and then describe the scope of our evaluation, the datasets used, the configuration of each approach, and the analysis methods employed.

\subsection{Research questions}
To guide our research, we formulated the following questions:

\begin{itemize}
\item \textbf{RQ1}: How does the performance of an LLM-based approach compare to a search-based approach for resolving merge conflicts?
\end{itemize}

The first research question aims to compare the performance of the two paradigms in terms of similarity to the expected resolution and the time required to generate a candidate resolution. We focus this initial comparison on conflicts from Java projects, the most common language in our datasets.

\begin{itemize}
\item \textbf{RQ2}: Does the performance difference between the LLM and search-based paradigms vary across different programming languages?
\end{itemize}

The second research question investigates whether the performance differences observed in RQ1 hold for conflicts arising in other programming languages, specifically C\#, JavaScript, and TypeScript. This assesses the language-dependent behavior of each paradigm.

\begin{itemize}
\item \textbf{RQ3}: How do the generalization capabilities of the two paradigms differ when evaluated on previously unseen datasets?
\end{itemize}

The third question focuses on assessing the generalizability of both paradigms. We analyze their performance when trained or tuned on one dataset and tested on another, revealing their adaptability to new and unfamiliar contexts.

\begin{itemize}
\item \textbf{RQ4}: What explains the performance differences between the LLM and search-based paradigms?
\end{itemize}

Lastly, the fourth research question seeks to understand the underlying factors that contribute to the superior performance of one paradigm over the other. This is achieved by qualitatively analyzing extreme cases where there is a significant performance difference between the two approaches.

\subsection{Scope of Evaluation: Combination-Based Conflicts}

It is important to note that the empirical evaluation in this study is scoped to conflicts that developers resolved exclusively by combining existing lines from the conflicting versions. This methodological choice is deliberate and justified by two primary factors.

First, this category of conflicts is highly prevalent and significant in practice. Large-scale empirical studies have shown that the vast majority of real-world merge conflicts are resolved without introducing entirely new code. For instance, \citet{ghiotto2020nature} found that in 87\% of analyzed conflicts, the final resolution was composed solely of lines already present in one of the conflicting versions. Therefore, focusing on combination-based resolutions allows our study to address a highly representative portion of the overall problem space.

Second, this scope enables a direct and fair comparison of the fundamental mechanisms of the GenAI and SBSE paradigms. The SBCR approach, by its combinatorial nature, is designed specifically to find optimal combinations of existing lines and cannot generate new code. While MergeGen is capable of generating novel content, evaluating it on ``new code'' conflicts against an approach that cannot do so would not yield a meaningful comparison of their core problem-solving strategies. By focusing on a task that is theoretically solvable by both, we create a level playing field. This allows us to rigorously assess their distinct strengths: MergeGen's learned, context-aware pattern matching versus SBCR's heuristic-driven optimization on a shared, well-defined problem.

While this scope does not test MergeGen's full generative potential, it provides the necessary focus to conduct a robust comparison of how each paradigm performs in the most common type of complex, non-trivial conflict resolution scenario.

\subsection{Datasets}

We used two datasets of real-world, combination-based merge conflicts collected from open-source projects.

$Dataset1$ was originally used by \citet{camposjunior2024composition} and initially contained 10,177 conflicting chunks from Java projects resolved by combination. After discarding chunks with empty content and reconstructing the base version for each conflict, the dataset comprises 6,269 conflicting chunks from 816 open-source Java projects hosted on GitHub.

$Dataset2$ was originally gathered by \citet{svyatkovskiy2022program} and used to evaluate MergeGen \citep{dong2023merge}. It initially contained 151,426 conflicts from projects in four programming languages. As our focus is on combination-based resolutions, we filtered this dataset accordingly, resulting in 47,363 conflicting chunks divided as follows: 20,728 from 2,203 Java projects ($Dataset2_{Java}$), 7,017 from 908 C\# projects ($Dataset2_{C\#}$), 14,151 from 3,205 JavaScript projects ($Dataset2_{JavaScript}$), and 5,467 from 1,083 TypeScript projects ($Dataset2_{TypeScript}$).

{\sloppy For all datasets ($Dataset1$, $Dataset2_{Java}$, $Dataset2_{C\#}$, $Dataset2_{JavaScript}$, $Dataset2_{TypeScript}$), we followed the approach adopted by \citet{dong2023merge} and \citet{svyatkovskiy2022program} to randomly split them into training, validation, and testing sets, using an approximate ratio of 8:1:1.\par}

To answer RQ1, we combine the results obtained from running both approaches on the test partition of $Dataset1$ and $Dataset2_{Java}$, since both datasets contain source code in Java. In this combination, we followed a conservative approach and excluded all chunks from both datasets that had the same commit hash to prevent duplicated data. Thus, in RQ1 we use the test partition of $Dataset_{Java}$, which contains 2,439 valid conflicting chunks.

\subsection{Experimental Setup}

\textbf{MergeGen Configuration}: MergeGen requires training on language-specific data. For our experiments, we trained a separate model for each language dataset. Due to computational constraints, we set the BPE token limits to 300 for the conflict input and 100 for the resolution output, which is smaller than the 500/200 token limits used in the original MergeGen study. Our experiments were conducted on a machine with an Intel i7-4790K CPU, 16GB of RAM, and an NVIDIA RTX 2080 GPU, a more modest setup than the one used in the original work. These differences may impact the performance and generalization capabilities of MergeGen in our evaluation.

\textbf{SBCR Configuration and Tuning}: SBCR is guided by three key parameters: \textit{neighbors\_per\_iteration}, \textit{max\_execution\_time}, and \textit{max\_stagnation\_iterations}. To find the optimal configuration for each dataset, we performed a systematic parameter tuning process on a random sample of 100 conflicts from each training set. We tested multiple values for each parameter (e.g., 1-9 neighbors, 10-40 seconds timeout, 5-20 stagnation iterations). The configuration that provided the best balance between solution similarity and execution time was selected and used for the final evaluation on the test set for that specific dataset. The tuning results are discussed in Section \ref{sec:tuning}.

\subsection{Evaluation Metrics and Analysis}

To compare the two paradigms, we used a combination of quantitative and qualitative analysis methods.

\textbf{Performance Metrics}: The primary metric for evaluating the quality of a generated resolution was \textbf{Similarity}, computed using the Gestalt pattern matching measure based on the Longest Common Subsequence (LCS) between the generated candidate and the developer's actual resolution. We also measured \textbf{Execution Time}. For SBCR, this includes the entire search process. For MergeGen, this measures only the generation time after receiving its specialized input, excluding the necessary pre-processing steps (e.g., tokenization, structural representation construction).

We opted for this measurement strategy to ensure a fair comparison focused on the core resolution mechanisms of each paradigm. SBCR operates directly on raw conflict text, while MergeGen requires a separate, and potentially time-consuming, pre-processing stage to create its structured input. By isolating only the resolution generation phase for both approaches, we assess the fundamental efficiency of the search-based algorithm versus the generative model. Consequently, the reported time for MergeGen represents a lower bound of its practical, end-to-end execution time, which would also include this pre-processing overhead.

\textbf{Statistical Analysis}: We first used the Shapiro-Wilk test to check if the data distributions were normal. As they were not, we used the non-parametric Wilcoxon Signed-Rank test \citep{wilcoxon1945individual} for paired comparisons between the approaches on the same set of conflicts. To measure the magnitude of the performance difference, we calculated the Common Language Effect Size (CLES) \citep{mcgraw1992common}, which represents the probability that a random result from one approach is superior to a result from the other.

\textbf{Qualitative Analysis for RQ4}: To understand the reasons behind performance differences, we calculated the metric $Sim_{SBCR-MergeGen}$ for each conflict. This metric represents the difference between the similarity score of the candidate generated by SBCR and the one generated by MergeGen ($Sim_{SBCR} - Sim_{MergeGen}$), allowing us to quantify the performance gap. Positive values indicate that SBCR performed better for a given conflict, while negative values indicate that MergeGen was superior. We then ranked all conflicts by this metric and selected the top 10 cases from each extreme of the distribution (highest positive and lowest negative values) for each of the five datasets, resulting in a total of \textbf{100 conflicts for manual analysis}. These extreme cases were inspected to identify recurring patterns and characteristics that could explain each paradigm's strengths and weaknesses.

\section{Results}
\label{sec:results}

In this section, we present the findings of our study, structured to address the key research questions. We report the results of the parameter tuning process for SBCR, followed by a series of experiments to evaluate and compare the performance of the SBSE and GenAI paradigms. We conclude by discussing potential threats to the validity of our findings.

\subsection{SBCR tuning}
\label{sec:tuning}

In this section, we present the tuning process for the SBCR approach, highlighting the selection of the best-performing configurations across various datasets. SBCR relies on three configurable parameters: the number of neighbors per iteration (\path{neighbors_per_iteration}), the maximum execution time (\path{max_execution_time}), and the stagnation limit (\path{max_stagnation_iterations}). These parameters were tuned to strike a balance between solution quality and efficiency. The configurations for each dataset were tested systematically to evaluate both the similarity of generated solutions to developer-provided resolutions and the required execution time. Each parameter combination's performance was measured by four key metrics: the Top-1 Similarity Count, Average Ranking, Average Similarity, and Average Time.

Table \ref{tbl:tuning} presents the metrics for the best-performing configurations of SBCR across each dataset, ordered by Average Similarity. We highlighted the configurations that were selected to be used in the experiments, based on their balance between average execution time and similarity to the resolution.

\begin{table}[]
\centering
\caption{Best-performing configurations of SBCR from the parameter tuning process across each dataset. Configurations are ordered by Average Similarity, and the highlighted rows were selected for the final experiments based on the optimal trade-off between similarity and execution time.}
\label{tbl:tuning}
\begin{tabular}{@{}lccrrr@{}}
\toprule
\multicolumn{1}{c}{\textbf{Dataset}} &
  \multicolumn{1}{c}{\textbf{Configuration}} &
  \multicolumn{1}{l}{
  \makecell{
    \textbf{Top-1} \\
    \textbf{Similarity} \\
    \textbf{Count}
  }
  } &
  \multicolumn{1}{c}{
  \makecell{
    \textbf{Average} \\
    \textbf{Ranking}
  }
  } &
  \multicolumn{1}{c}{
  \makecell{
    \textbf{Average} \\
    \textbf{Similarity}
  }
  } &
  \multicolumn{1}{c}{
  \makecell{
    \textbf{Average} \\
    \textbf{Time}
  }
  }
  \\ \midrule
 &
  7, 10, 30 &
  46/100 &
  9.52 &
  0.827 &
  12.28 \\
 &
  \cellcolor[HTML]{C0C0C0}5, 15, 10 &
  \cellcolor[HTML]{C0C0C0}45/100 &
  \cellcolor[HTML]{C0C0C0}10.33 &
  \cellcolor[HTML]{C0C0C0}0.827 &
  \cellcolor[HTML]{C0C0C0}5.96 \\
\multirow{-3}{*}{$Dataset1$} &
  7,  15, 40 &
  46/100 &
  10.14 &
  0.821 &
  14.79 \\ \midrule
 &
  \cellcolor[HTML]{C0C0C0}5, 10, 20 &
  \cellcolor[HTML]{C0C0C0}41/100 &
  \cellcolor[HTML]{C0C0C0}16.18 &
  \cellcolor[HTML]{C0C0C0}0.827 &
  \cellcolor[HTML]{C0C0C0}4.44 \\
 &
  9, 10, 20 &
  39/100 &
  11.39 &
  0.826 &
  18.41 \\
\multirow{-3}{*}{$Dataset2_{Java}$} &
  9, 15, 40 &
  39/100 &
  11.66 &
  0.825 &
  22.93 \\ \midrule
 &
  \cellcolor[HTML]{C0C0C0}3, 10, 40 &
  \cellcolor[HTML]{C0C0C0}36/100 &
  \cellcolor[HTML]{C0C0C0}15.15 &
  \cellcolor[HTML]{C0C0C0}0.791 &
  \cellcolor[HTML]{C0C0C0}2.32 \\
 &
  5, 10, 20 &
  38/100 &
  13.90 &
  0.790 &
  4.69 \\
\multirow{-3}{*}{$Dataset2_{C\#}$} &
  9, 10, 10 &
  38/100 &
  11.73 &
  0.786 &
  12.27 \\ \midrule
 &
  9, 5, 40 &
  34/100 &
  16.91 &
  0.833 &
  14.48 \\
 &
  9, 20, 40 &
  36/100 &
  14.01 &
  0.824 &
  22.54 \\
\multirow{-3}{*}{$Dataset2_{JavaScript}$} &
  \cellcolor[HTML]{C0C0C0}9, 15, 20 &
  \cellcolor[HTML]{C0C0C0}36/100 &
  \cellcolor[HTML]{C0C0C0}15.86 &
  \cellcolor[HTML]{C0C0C0}0.818 &
  \cellcolor[HTML]{C0C0C0}13.74 \\ \midrule
 &
  \cellcolor[HTML]{C0C0C0}9, 15, 10 &
  \cellcolor[HTML]{C0C0C0}20/100 &
  \cellcolor[HTML]{C0C0C0}14.49 &
  \cellcolor[HTML]{C0C0C0}0.802 &
  \cellcolor[HTML]{C0C0C0}8.16 \\
 &
  7, 20, 30 &
  20/100 &
  16.77 &
  0.801 &
  12.11 \\
\multirow{-3}{*}{$Dataset2_{TypeScript}$} &
  9, 20, 30 &
  18/100 &
  15.24 &
  0.798 &
  17.19 \\ \bottomrule
\end{tabular}
\end{table}

\subsection{RQ1: How does the performance of an LLM-based approach compare to a search-based approach for resolving merge conflicts?}

To answer this research question, we compare the performance of SBCR and MergeGen on the combined Java test partition ($Dataset_{Java}$), which contains 2,439 conflicts.

\begin{figure}[htbp]
    \centering
    \includegraphics[width=0.8\linewidth]{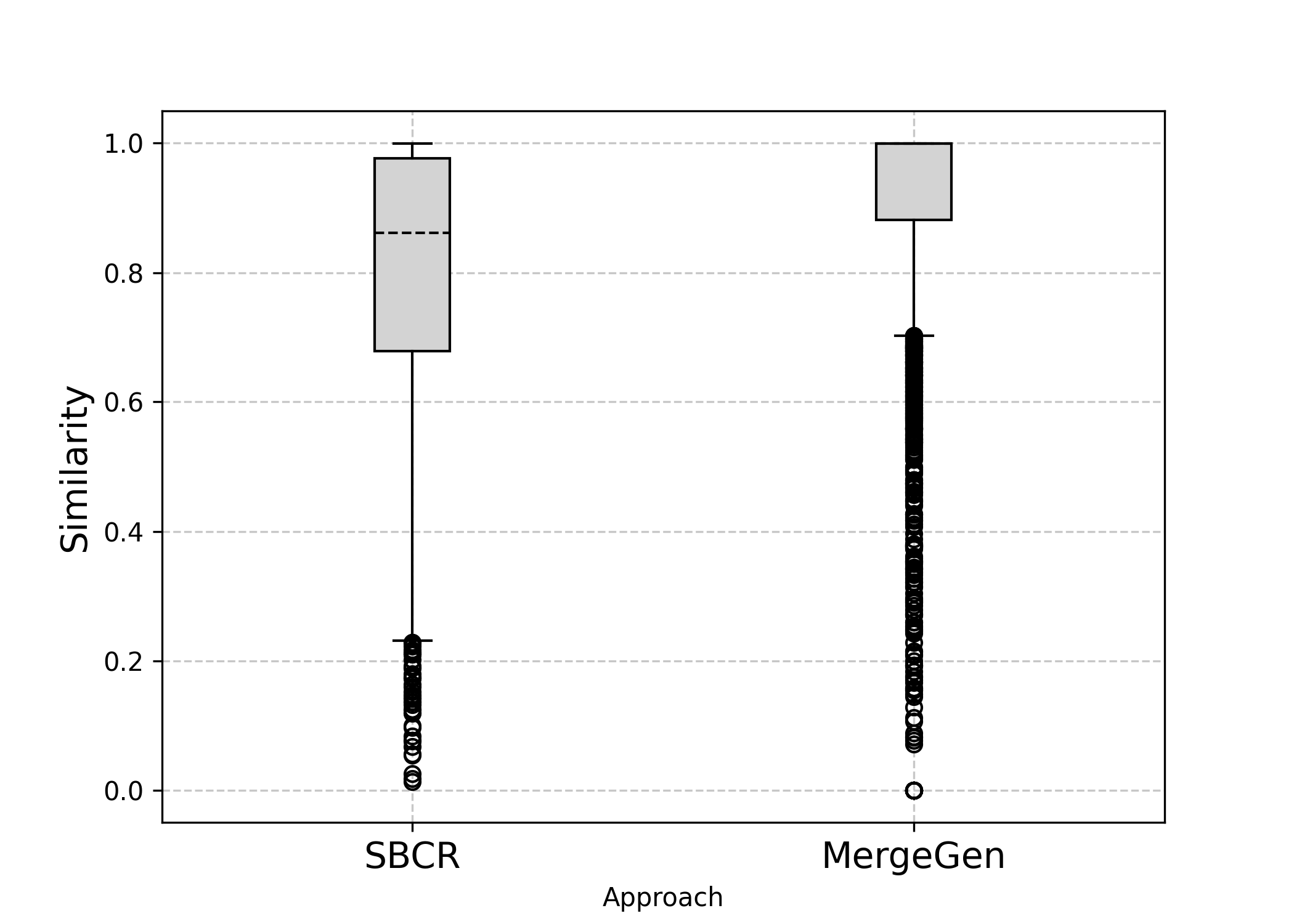}
    \caption{Boxplots for the similarities between the resolution adopted by the developers and the conflict resolution candidates generated by SBCR and MergeGen for Java projects.}
    \label{fig:rq1_similarity_boxplot}
\end{figure}

The average similarity of candidates generated by MergeGen to the expected resolution is 90.0\% (std. dev. 19.3\%), while for SBCR it is 79.9\% (std. dev. 20.6\%) . As shown in Figure \ref{fig:rq1_similarity_boxplot}, MergeGen's performance is notably strong, achieving a median similarity of 100\% and resolving 55\% of conflicts with a perfect match. In comparison, SBCR achieves a median similarity of 86.1\%, with perfect matches in 19.6\% of cases. A Wilcoxon Signed-Rank test \citep{wilcoxon1945individual} confirmed that this difference is statistically significant ($p<0.001$). To measure the effect size, we calculated the Common Language Effect Size (CLES). The computed CLES is 0.294, meaning there is a 29.4\% probability that SBCR will generate a more similar resolution than MergeGen for a randomly chosen conflict, while MergeGen has a 70.6\% probability of being superior.

\begin{figure}[htbp]
    \centering
    \includegraphics[width=0.8\linewidth]{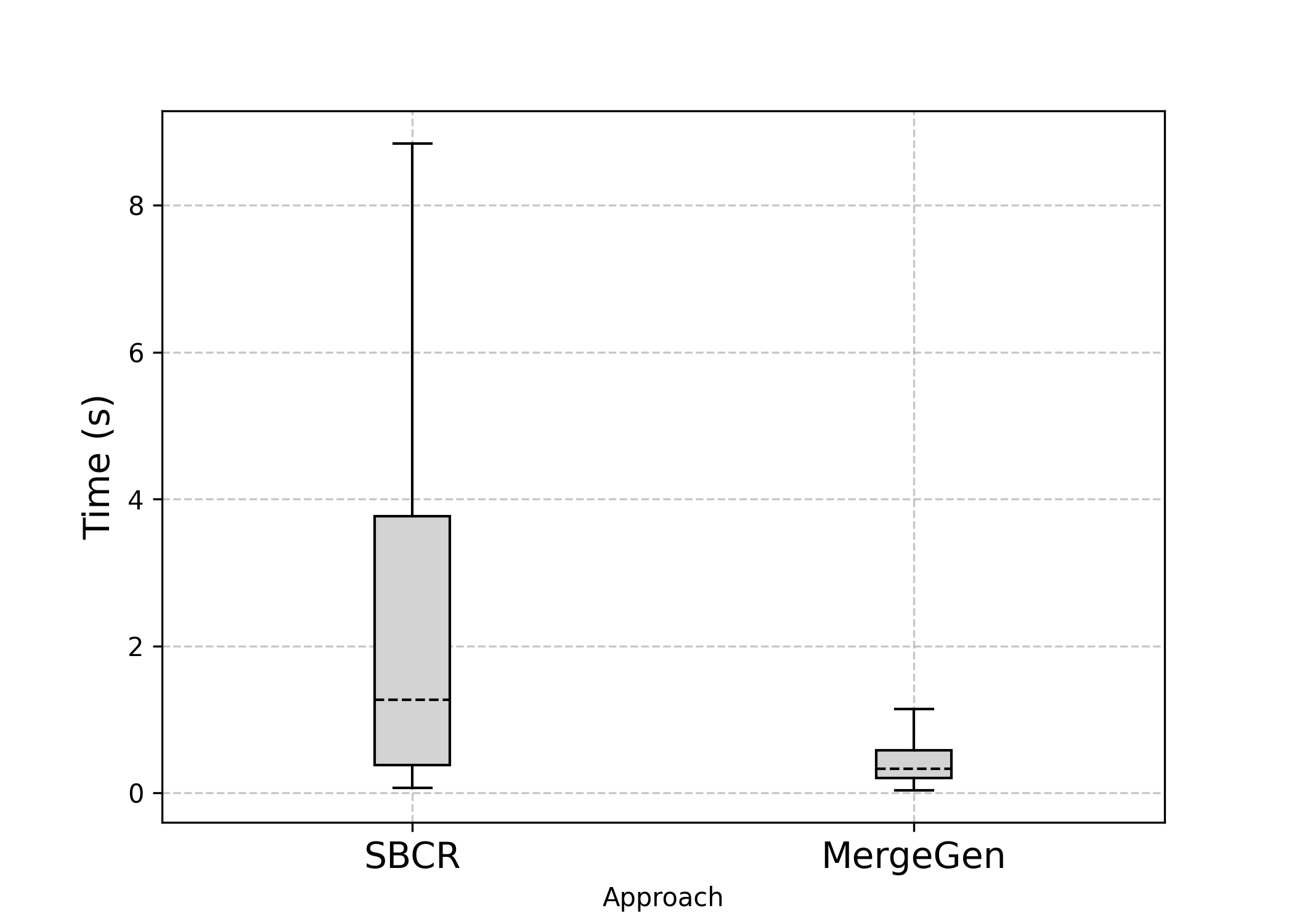}
    \caption{Boxplots for the time in seconds each approach takes to generate a candidate resolution for a conflicting chunk.}
    \label{fig:rq1_time_boxplot}
\end{figure}

Regarding execution time (Figure \ref{fig:rq1_time_boxplot}), MergeGen is demonstrably faster, with a median time of 0.3 seconds, compared to SBCR's 1.3 seconds. We also performed a hypothesis test to confirm this observation. A Shapiro-Wilk test showed that the time distributions for both approaches are not normal ($p < 0.001$). Therefore, we used a Wilcoxon Signed-Rank test, which revealed a statistically significant difference between the two approaches ($p < 0.001$). The Common Language Effect Size (CLES) was 0.759, indicating that for a randomly selected conflict, there is a 75.9\% probability that SBCR's execution time will be longer than MergeGen's, confirming the latter's substantial speed advantage.

\begin{center}
\noindent\fbox{%
    \parbox{0.95\columnwidth}{
        \textbf{Finding 1}: For Java conflicts, the GenAI paradigm (MergeGen) produces more accurate resolutions than the SBSE paradigm (SBCR), with a higher median similarity (100\% vs. 86.1\%) and a greater number of perfect matches (55\% vs. 19.6\%). The difference is statistically significant, with MergeGen generating a more similar resolution in 70.6\% of cases (CLES = 0.294). MergeGen is also significantly faster (median 0.3s vs 1.3s, $p < 0.001$), with a 75.9\% probability of being faster on any given conflict (CLES = 0.759).
    }
}
\end{center}

\subsection{RQ2: Does the performance difference between the paradigms vary across different programming languages?}

\begin{figure}[htbp]
    \centering
    \includegraphics[width=\linewidth]{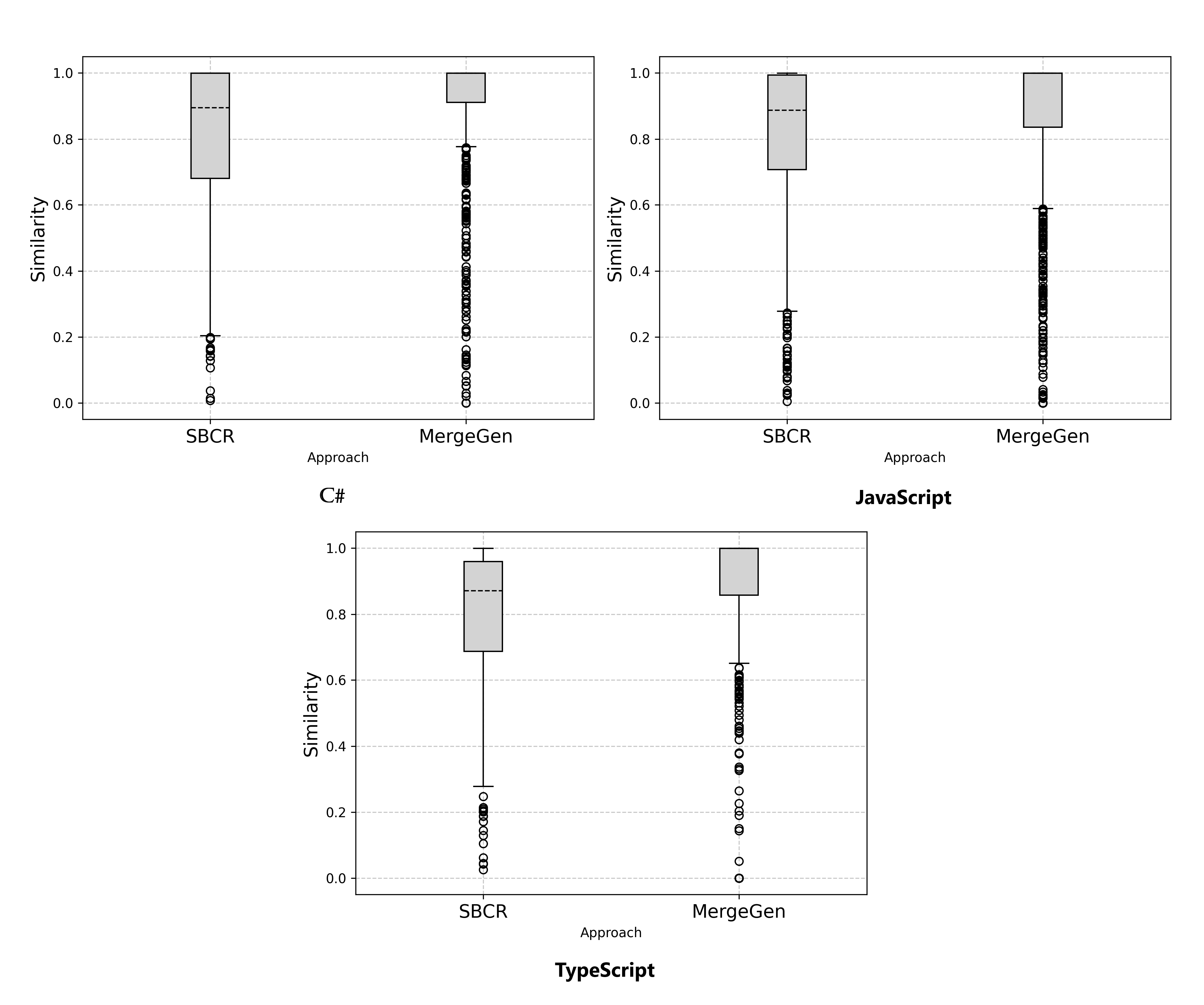}
    \caption{Boxplots for the similarities between the resolution adopted by the developers and the conflict resolution candidates generated by SBCR and MergeGen for C\#, JavaScript, and TypeScript projects, respectively.}
    \label{fig:rq2_similarity_boxplots}
\end{figure}

To answer RQ2, we investigate the performance of both paradigms on conflicts from C\#, JavaScript, and TypeScript projects. As shown by the boxplots in Figure \ref{fig:rq2_similarity_boxplots} and the descriptive statistics in Table \ref{tbl:rq2_statistics_table}, the GenAI paradigm (MergeGen) consistently achieves higher similarity scores than the SBSE paradigm (SBCR) across all three languages. MergeGen reaches a median similarity of 1.00 in all cases, while SBCR's medians are 0.89 for C\# and JavaScript, and 0.87 for TypeScript.

\begin{table}[ht]
\centering
\caption{Statistics for the similarity results obtained when generating resolution candidates using each approach for C\#, JavaScript, and TypeScript projects.}
\label{tbl:rq2_statistics_table}
\begin{tabular}{@{}lcccccc@{}}
\toprule
\multirow{2}{*}{
\makecell{
\textbf{Programming} \\
\textbf{language}
}
} & \multicolumn{3}{c}{\textbf{SBCR}} & \multicolumn{3}{c}{\textbf{MergeGen}} \\ \cmidrule(lr){2-4} \cmidrule(lr){5-7}
                      & Mean & Std. dev. & Median & Mean & Std. dev. & Median \\ \midrule
C\#                   & 0.80 & 0.24      & 0.89   & 0.89 & 0.23      & 1.00   \\
JavaScript            & 0.81 & 0.21      & 0.89   & 0.88 & 0.22      & 1.00    \\
TypeScript            & 0.79 & 0.22      & 0.87   & 0.87 & 0.25      & 1.00    \\ \bottomrule
\end{tabular}
\end{table}

To determine if these observed differences are statistically significant, we first performed a Shapiro-Wilk test to assess the normality of the data distributions for each approach and language. In all cases, the test indicated that the data is not normally distributed ($p < 0.001$). Consequently, we used the non-parametric Wilcoxon Signed-Rank test for our paired comparisons. The test confirmed that the differences in similarity scores between MergeGen and SBCR are highly significant for all three languages (C\#, JavaScript, and TypeScript), with $p < 0.001$ in each case.

To understand the magnitude of this difference, we calculated the Common Language Effect Size (CLES). The results indicate that MergeGen's advantage is substantial: SBCR has only a 33.4\% chance of outperforming MergeGen in a random C\# conflict, a 34.3\% chance in JavaScript, and a 29.3\% chance in TypeScript.

While MergeGen's performance is remarkably consistent, SBCR shows slightly lower performance for TypeScript, suggesting its text-based approach may be more sensitive to language-specific characteristics. In contrast, MergeGen, which leverages a language model-based approach, appears to generalize more effectively across programming languages. This robustness could be attributed to its richer, conflict-aware representation of the code.

\begin{center}
\noindent\fbox{%
    \parbox{0.95\columnwidth}{
        \textbf{Finding 2}: The GenAI paradigm (MergeGen) consistently outperforms the SBSE paradigm (SBCR) across C\#, JavaScript, and TypeScript, achieving a perfect median similarity of 1.00 in all languages. The differences are statistically significant, with CLES values confirming MergeGen's advantage in approximately 66-71\% of cases, depending on the language.
    }
}
\end{center}

\subsection{RQ3: How do the generalization capabilities of the two paradigms differ?}

To assess generalization, we analyze how well SBCR and MergeGen perform when trained or tuned on one dataset and then evaluated on another. We conduct this analysis in two stages: first testing on $Dataset1$ and then on $Dataset2_{Java}$.

\subsubsection{Evaluation on $Dataset1$}

First, we evaluated both approaches on $Dataset1$. This stage allows us to compare two distinct scenarios for generalization. The first is a \textbf{within-dataset} evaluation, where the approaches are tuned/trained and tested on different partitions of the same dataset ($Dataset1$). In this scenario, while the test data is unseen, the models have been exposed to other conflicts from the same set of projects, potentially learning project-specific styles and APIs. The second is a \textbf{cross-dataset} evaluation, where the approaches are tuned/trained on an entirely different set of projects ($Dataset2_{Java}$) and tested on $Dataset1$. This is a more challenging test of generalization, as the models have no prior exposure to the coding styles or context of the test projects.

Table \ref{tbl:sbcr-rq3_descriptive_dataset1} presents the descriptive statistics for this stage. Notably, the performance of SBCR is nearly identical in both the within-dataset and cross-dataset scenarios, with mean similarities of 0.809 and 0.810, respectively. This demonstrates its strong ability to generalize to completely unfamiliar projects.

\begin{table}[ht]
\centering
\caption{Descriptive statistics for similarities obtained when using each approach tuned or trained on $Dataset1$ and $Dataset2_{Java}$ and tested on $Dataset1$ conflicts.}
\label{tbl:sbcr-rq3_descriptive_dataset1}
\begin{tabular}{@{}lcccccc@{}}
\toprule
\multirow{2}{*}{\textbf{Tuning / Training}} & \multicolumn{3}{c}{\textbf{SBCR}} & \multicolumn{3}{c}{\textbf{MergeGen}} \\
\cmidrule(lr){2-4} \cmidrule(lr){5-7}
 & Mean & Std. dev. & Median & Mean & Std. dev. & Median \\
\midrule
$Dataset1$        & 0.809 & 0.200 & 0.865 & 0.923 & 0.153 & 1.00 \\
$Dataset2_{Java}$ & 0.810 & 0.201 & 0.865 & 0.902 & 0.194 & 1.00 \\
\bottomrule
\end{tabular}
\end{table}

\begin{figure}[htbp]
    \centering
    \includegraphics[width=\linewidth]{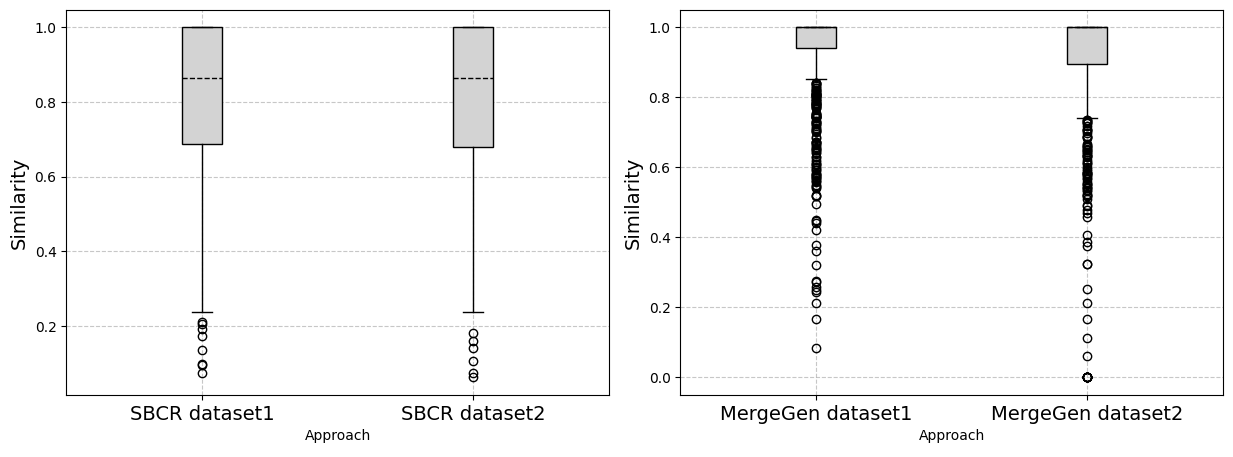}
    \caption[Box plots comparing SBCR and MergeGen similarity scores when tuned on $Dataset1$ and $Dataset2_{Java}$, and tested on $Dataset1$.]{Box plots comparing the similarity of candidates generated by SBCR and MergeGen, both tuned on $Dataset1$ and $Dataset2_{Java}$, and tested on $Dataset1$. The left subplot shows the comparison between SBCR tuned on both datasets, while the right subplot shows the comparison between MergeGen trained on both datasets.}
    \label{fig:sbcr-rq3_each_approach_dataset1}
\end{figure}

The boxplots in Figure \ref{fig:sbcr-rq3_each_approach_dataset1} visually confirm this stability for SBCR (left subplot) and show a slight variation for MergeGen (right subplot).

The statistical tests, detailed in Table \ref{tbl:sbcr-rq3_significance_dataset1}, support these observations. For SBCR, a Wilcoxon p-value of 0.953 and a CLES of 0.497 indicate no practical or statistically significant difference. For MergeGen, however, training on a different dataset ($Dataset2_{Java}$) resulted in a slight but statistically significant performance drop, as supported by a p-value of 0.022 and a CLES of 0.525, indicating a small advantage when trained on $Dataset1$.

\begin{table}[ht]
\centering
\caption{Statistical significance tests results for the approaches' comparisons using different training sets and evaluated on $Dataset1$.}
\label{tbl:sbcr-rq3_significance_dataset1}
\begin{adjustbox}{max width=\linewidth}
\begin{tabular}{@{}ll llcr@{}}
\toprule
\textbf{Approach} & \textbf{Tuning/Training} & \textbf{Approach} & \textbf{Tuning/Training} &
\makecell{\textbf{Wilcoxon's}  \\ \textbf{p-value}}
 & \textbf{CLES} \\
\midrule
SBCR     & $Dataset1$        & SBCR     & $Dataset2_{Java}$ & 0.953                   & 0.497 \\
MergeGen & $Dataset1$        & MergeGen & $Dataset2_{Java}$ & 0.022                   & 0.525 \\
SBCR     & $Dataset2_{Java}$ & MergeGen & $Dataset2_{Java}$ & $1.617 \times 10^{-25}$ & 0.322 \\
\bottomrule
\end{tabular}
\end{adjustbox}
\end{table}

\begin{figure}[htbp]
    \centering
    \includegraphics[width=0.8\linewidth]{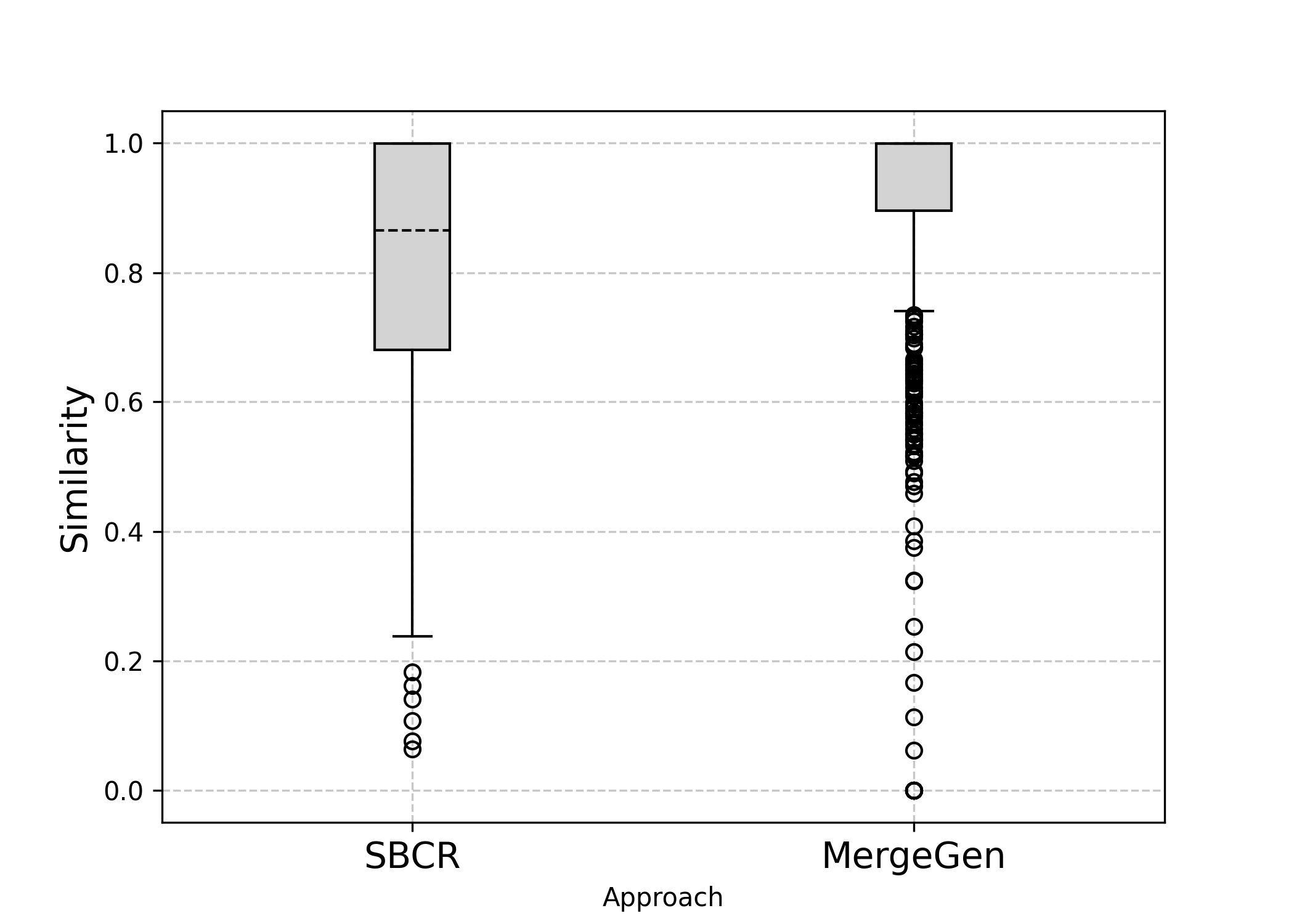}
    \caption[Box plot comparing SBCR and MergeGen similarity scores when trained on $Dataset2_{Java}$ and tested on $Dataset1$]{Box plot comparing the similarity of candidates generated by SBCR and MergeGen, both trained on $Dataset2_{Java}$, and tested on $Dataset1$. This figure highlights the relative performance of both approaches on $Dataset1$ conflicts, when trained on a different dataset.}
    \label{fig:sbcr-rq3_versus_dataset1}
\end{figure}

Despite this sensitivity, MergeGen still significantly outperforms SBCR in the cross-dataset scenario. Figure \ref{fig:sbcr-rq3_versus_dataset1} compares both approaches when trained/tuned on $Dataset2_{Java}$ and tested on $Dataset1$. The difference is statistically significant ($p < 0.001$) with a CLES of 0.322, meaning SBCR is likely to produce a better resolution than MergeGen in 32.2\% of cases in this scenario, as also detailed in Table \ref{tbl:sbcr-rq3_significance_dataset1}.

\subsubsection{Evaluation on $Dataset2_{Java}$}

Second, we repeated the evaluation on the $Dataset2_{Java}$ test set. The results, summarized in Table \ref{tbl:sbcr-rq3_descriptive_dataset2}, show a similar pattern. SBCR again demonstrates robust generalization, with no statistically significant performance difference when tuned on $Dataset1$ versus $Dataset2_{Java}$.

\begin{table}[ht]
\centering
\caption{Descriptive statistics for similarities obtained when using each approach tuned on $Dataset1$ and $Dataset2_{Java}$ and tested on $Dataset2_{Java}$ conflicts.}
\label{tbl:sbcr-rq3_descriptive_dataset2}
\begin{tabular}{@{}lcccccc@{}}
\toprule
\multirow{2}{*}{\textbf{Tuning / Training}} & \multicolumn{3}{c}{\textbf{SBCR}} & \multicolumn{3}{c}{\textbf{MergeGen}} \\
\cmidrule(lr){2-4} \cmidrule(lr){5-7}
 & Mean & Std. dev. & Median & Mean & Std. dev. & Median \\
\midrule
$Dataset1$        & 0.799 & 0.206 & 0.860 & 0.894 & 0.190 & 1.000 \\
$Dataset2_{Java}$ & 0.795 & 0.207 & 0.858 & 0.893 & 0.204 & 1.000 \\
\bottomrule
\end{tabular}
\end{table}

\begin{figure}[htbp]
    \centering
    \includegraphics[width=\linewidth]{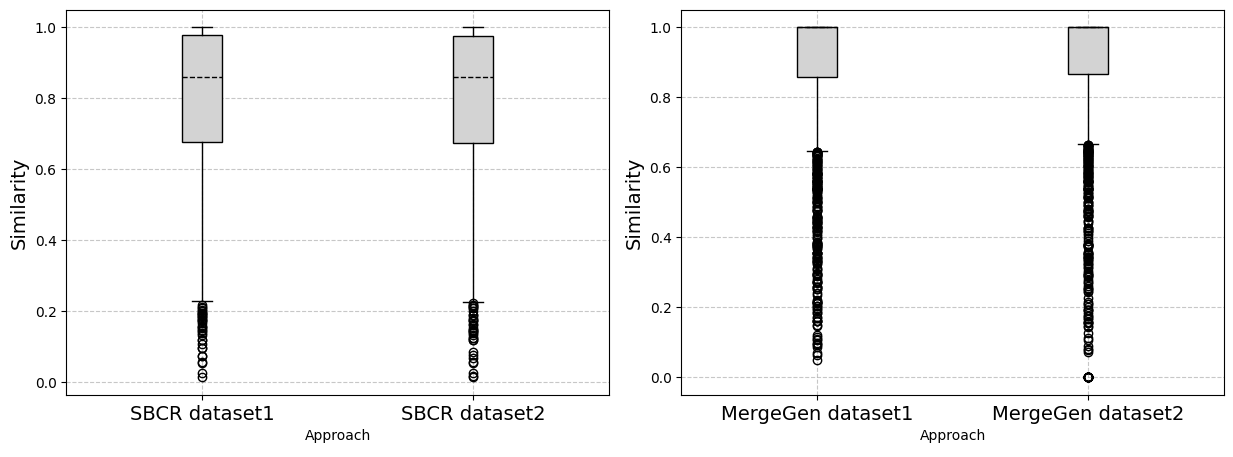}
    \caption[Box plots comparing SBCR and MergeGen similarity scores when tuned on $Dataset1$ and $Dataset2_{Java}$, and tested on $Dataset2_{Java}$]{Box plots comparing the similarity of candidates generated by SBCR and MergeGen, both tuned on $Dataset1$ and $Dataset2_{Java}$, and tested on $Dataset2_{Java}$. The left subplot shows the comparison between SBCR tuned on both datasets, while the right subplot shows the comparison between MergeGen trained on both datasets.}
    \label{fig:sbcr-rq3_each_approach_dataset2}
\end{figure}

The consistency of SBCR is also visible in Figure \ref{fig:sbcr-rq3_each_approach_dataset2} (left). In contrast, MergeGen again showed a statistically significant, though minor, sensitivity to the training data, as detailed in Table \ref{tbl:sbcr-rq3_significance_dataset2} ($p = 0.021$ and $CLES = 0.490$).

\begin{table}[ht]
\centering
\caption{Statistical significance tests results for the approaches' comparisons using different training sets and evaluated on $Dataset2_{Java}$.}
\label{tbl:sbcr-rq3_significance_dataset2}
\begin{adjustbox}{max width=\linewidth}
\begin{tabular}{@{}ll llcr@{}}
\toprule
\textbf{Approach} & \textbf{Tuning/Training} & \textbf{Approach} & \textbf{Tuning/Training} &
\makecell{\textbf{Wilcoxon's}  \\ \textbf{p-value}}
& \textbf{CLES} \\
\midrule
SBCR     & $Dataset1$ & SBCR     & $Dataset2_{Java}$ & 0.230                   & 0.506 \\
MergeGen & $Dataset1$ & MergeGen & $Dataset2_{Java}$ & 0.021                   & 0.490 \\
SBCR     & $Dataset1$ & MergeGen & $Dataset1$        & $2.45 \times 10^{-80}$ & 0.308 \\
\bottomrule
\end{tabular}
\end{adjustbox}
\end{table}

\begin{figure}[htbp]
    \centering
    \includegraphics[width=0.8\linewidth]{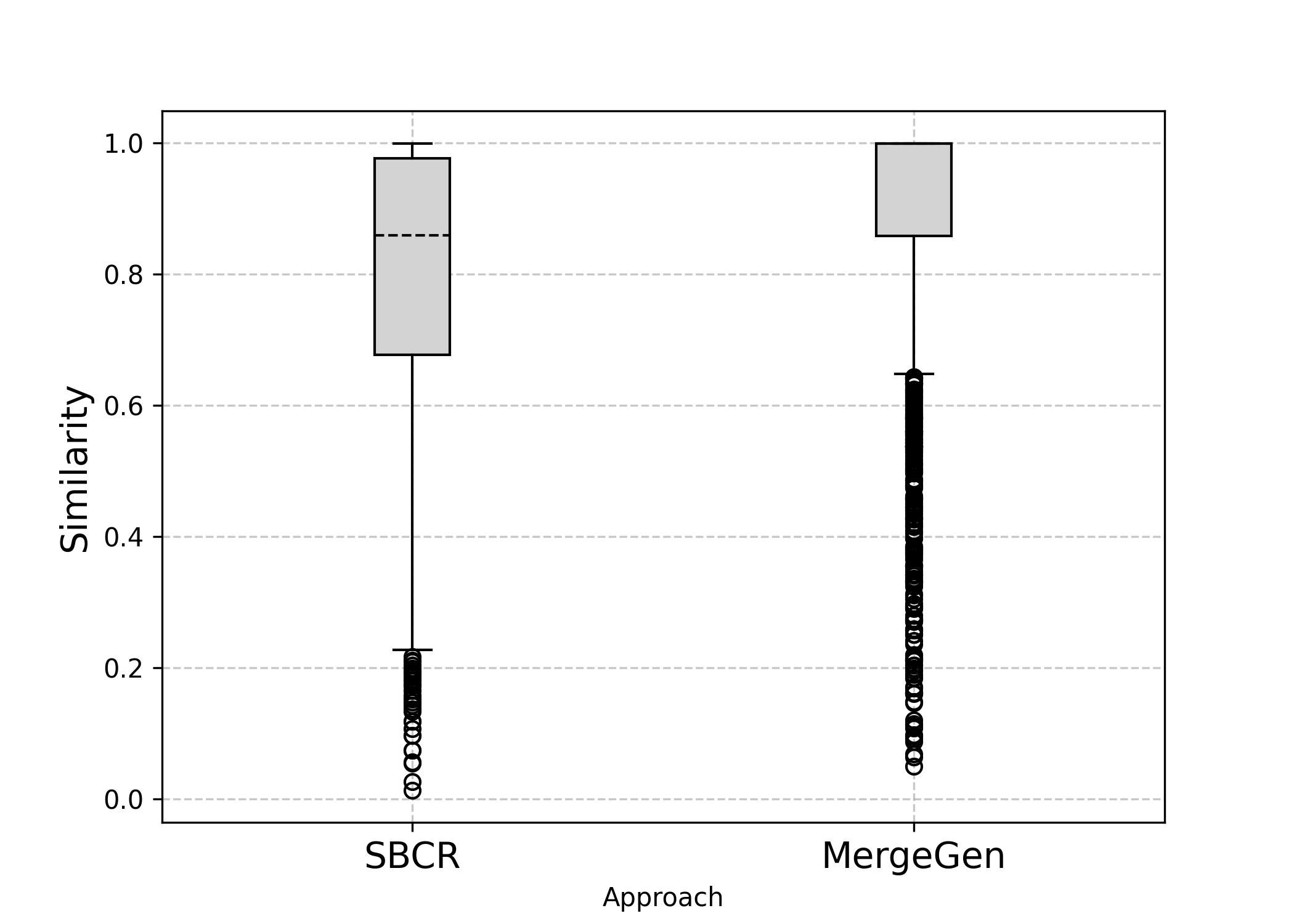}
    \caption[Box plot comparing SBCR and MergeGen similarity scores when trained on $Dataset1$ and tested on $Dataset2_{Java}$.]{Box plot comparing the similarity of candidates generated by SBCR and MergeGen, both trained on $Dataset1$, and tested on $Dataset2_{Java}$. This figure illustrates the direct comparison of both approaches when trained on $Dataset1$ and evaluated on $Dataset2_{Java}$.}
    \label{fig:sbcr-rq3_versus_dataset2}
\end{figure}

When directly compared on the $Dataset2_{Java}$ test set (with both approaches trained/tuned on $Dataset1$), MergeGen again significantly outperforms SBCR, as seen in Figure \ref{fig:sbcr-rq3_versus_dataset2}. This difference is confirmed to be significant ($p < 0.001$) with a CLES of 0.308, as shown in Table \ref{tbl:sbcr-rq3_significance_dataset2}.

\vspace{1em}

\begin{center}
\noindent\fbox{%
    \parbox{0.95\columnwidth}{
        \textbf{Finding 3}: The SBSE paradigm (SBCR) demonstrates superior generalization, with its performance remaining stable across different datasets. The GenAI paradigm (MergeGen) is slightly sensitive to its training data but consistently achieves higher median similarity scores, outperforming SBCR even in cross-dataset evaluations.
    }
}
\end{center}

\subsection{RQ4: What explains the performance differences between the two paradigms?}

To understand the factors driving the performance differences, we analyzed each conflict using the $Sim_{SBCR-MergeGen}$ metric, which subtracts MergeGen's similarity score from SBCR's. This allows us to quantify which paradigm performed better for each specific case.

First, we summarize the overall distribution of wins, losses, and ties in Table \ref{tbl:sbcr-rq4_qty_summary}.

\begin{table}[ht]
\centering
\caption{Summary of chunk performance comparison between SBCR and MergeGen for each dataset in terms of similarity to the resolution.}
\label{tbl:sbcr-rq4_qty_summary}
\begin{adjustbox}{max width=\linewidth}
\begin{tabular}{@{}lrrrr@{}}
\toprule
\textbf{Dataset}                 & $\text{SBCR} > \text{MergeGen}$ & $\text{MergeGen} > \text{SBCR}$ & $\text{SBCR} = \text{MergeGen}$ & \textbf{Total} \\
\midrule
$Dataset1$            & 131  & 380  & 117 & 628  \\
$Dataset2_{Java}$       & 430  & 1268 & 269 & 1967 \\
$Dataset2_{C\#}$     & 148  & 318  & 143 & 609  \\
$Dataset2_{JavaScript}$ & 331  & 725  & 239 & 1295 \\
$Dataset2_{TypeScript}$ & 104  & 327  & 76  & 507  \\
\bottomrule
\end{tabular}
\end{adjustbox}
\end{table}

\begin{figure}[htbp]
    \centering
    \includegraphics[width=0.8\linewidth]{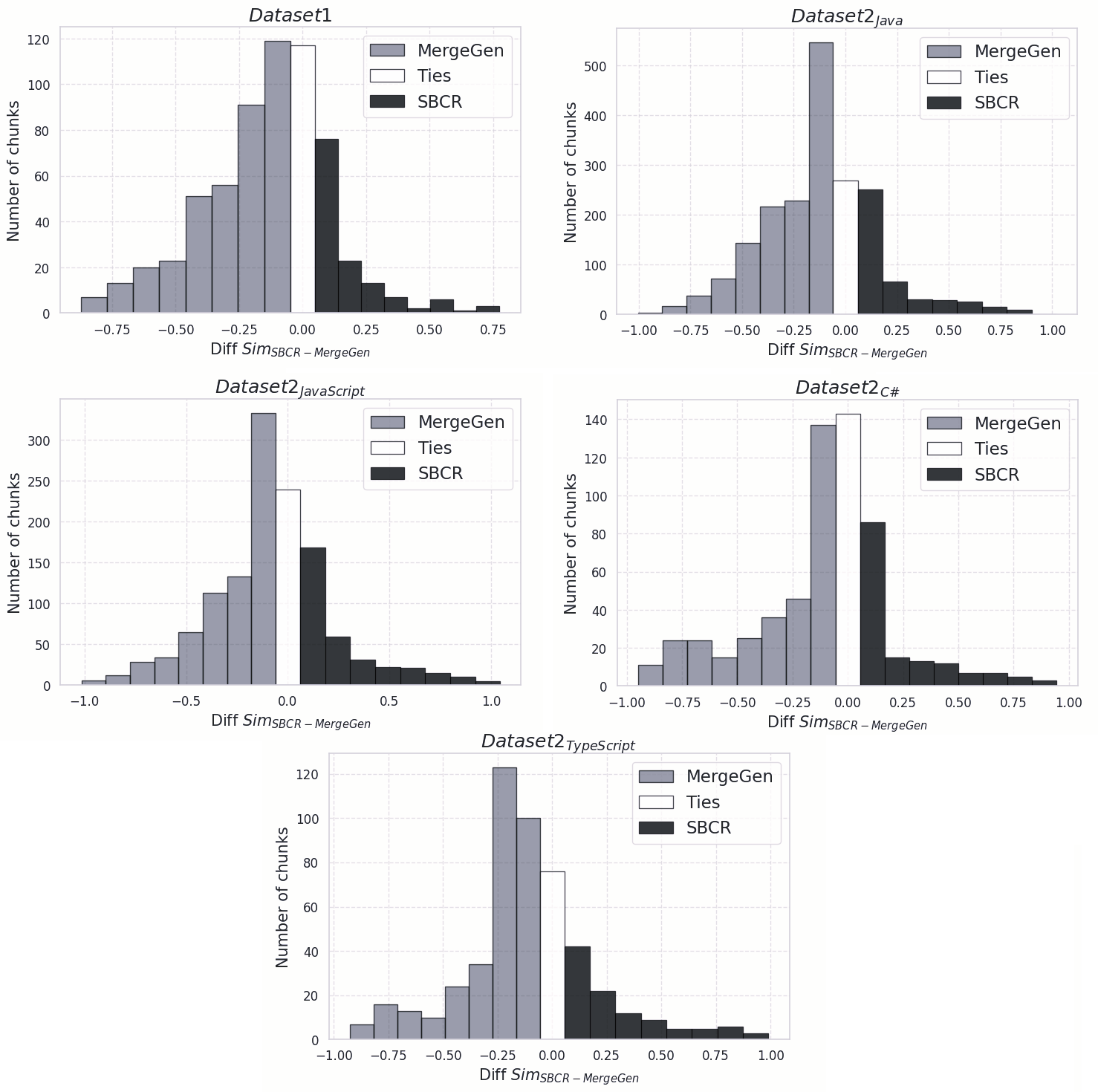}
    \caption[Distribution of $Diff_{SBCR-MergeGen}$ values across all datasets.]{Distribution of $Diff_{SBCR-MergeGen}$ values across all datasets. Negative values (gray) indicate better performance by MergeGen, positive values (black) indicate better performance by SBCR, and zero (white bars) indicate ties.}
    \label{fig:sbcr-rq4_diff_distribution}
\end{figure}

According to Table \ref{tbl:sbcr-rq4_qty_summary}, MergeGen outperformed SBCR in the majority of cases across all datasets. For instance, in $Dataset2_{Java}$, MergeGen was superior in 1,268 (64.5\%) of the conflicts, while SBCR was superior in 430 (21.9\%). The histograms in Figure \ref{fig:sbcr-rq4_diff_distribution} visualize this distribution.

The histograms in Figure \ref{fig:sbcr-rq4_diff_distribution} confirm that while many cases cluster around zero (indicating comparable performance), the distribution is skewed to the left, favoring MergeGen. To understand the reasons behind the performance differences shown in Figure \ref{fig:sbcr-rq4_diff_distribution}, we proceeded with the qualitative analysis detailed in our methodology. This involved a manual inspection of a balanced set of 100 conflicts drawn from the extremes of the performance distribution---50 cases where SBCR was superior and 50 where MergeGen was superior, spread evenly across all five datasets. Our inspection of these cases revealed clear and recurring patterns that define the contextual strengths and weaknesses of each paradigm. The following sections present the categorized findings from this manual analysis.

\subsubsection{Qualitative Analysis: When MergeGen Excels}

Our manual analysis revealed that MergeGen's primary strength when compared to SBCR lies in handling conflicts with significant structural imbalance. The specific scenarios where MergeGen was much better than SBCR are categorized in Table \ref{tbl:sbcr-rq4_mergegen_better_sbcr}.

\begin{table}[ht]
\centering
\caption{Situations where MergeGen performed much better than SBCR, categorized by dataset.}
\label{tbl:sbcr-rq4_mergegen_better_sbcr}
\begin{adjustbox}{max width=\linewidth}
\begin{tabular}{@{}p{5.5cm} rrrrr r@{}}
\toprule
\textbf{Situation} & \textbf{D1} & \textbf{D2\textsubscript{Java}} & \textbf{D2\textsubscript{C\#}} & \textbf{D2\textsubscript{JS}} & \textbf{D2\textsubscript{TS}} & \textbf{Total} \\
\midrule
Use of few lines from the larger version & 1 & 2 & 4 & 4 & 2 & 13 \\
Use of almost all lines from both versions, where one is bigger & 2 & 4 & 5 & 0 & 1 & 12 \\
Whitespace or removed content & 4 & 2 & 1 & 2 & 0 & 9 \\
Few lines from smaller version, many from larger version & 2 & 0 & 0 & 0 & 4 & 6 \\
Use of a few lines from smaller version & 1 & 2 & 0 & 0 & 0 & 3 \\
Use of a few lines from both versions & 0 & 0 & 0 & 1 & 2 & 3 \\
Chunk large compared to resolution & 0 & 0 & 0 & 2 & 1 & 3 \\
Malformed chunk & 0 & 0 & 0 & 1 & 0 & 1 \\
\midrule
\textbf{Total} & \textbf{10} & \textbf{10} & \textbf{10} & \textbf{10} & \textbf{10} & \textbf{50} \\
\bottomrule
\end{tabular}
\end{adjustbox}
\vspace{0.5em}\\
{\footnotesize \textit{Note:} D1 = $Dataset1$; D2\textsubscript{Java} = $Dataset2_{Java}$; D2\textsubscript{C\#} = $Dataset2_{C\#}$; D2\textsubscript{JS} = $Dataset2_{JavaScript}$; D2\textsubscript{TS} = $Dataset2_{TypeScript}$.}

\end{table}

As categorized in Table \ref{tbl:sbcr-rq4_mergegen_better_sbcr}, this imbalance manifested in several ways. In 50\% of the analyzed cases (25 out of 50), the resolution involved selecting either a small subset of lines from a much larger version or nearly all lines from both versions despite a large size disparity. In another 18\% of cases, the imbalance was caused by one version removing content that the other modified. In these scenarios, SBCR's evaluation function, which seeks to balance contributions, was ineffective. In contrast, MergeGen's learned contextual understanding allowed it to generate the correct, unbalanced resolution, as exemplified in Listing \ref{lst:sbcr-example_chunk_overfitting}.

\begin{lstlisting}[caption={Example of a conflicting chunk where MergeGen correctly inferred the resolution by keeping only the \texttt{@Override} annotation. SBCR's balancing strategy led it to preserve most lines from the larger version, resulting in an incorrect resolution.}, label={lst:sbcr-example_chunk_overfitting}, float]
<<<<<<<
=======

    /**
     * For testing
     */
    ChronicleMapBuilder<K, V> forceReplicatedImpl() {
        this.forceReplicatedImpl = true;
        return this;
    }

    @Override
>>>>>>>
\end{lstlisting}

Listing \ref{lst:sbcr-example_chunk_overfitting} displays a conflicting chunk (file \texttt{ChronicleMapBuilder.java} in merge \href{https://github.com/OpenHFT/Chronicle-Map/commit/9d8e848#diff-7dc394e89e0b32fd71fad0c5420a00dc2565b0ea84ffa1e3228ce84348db247c}{9d8e848} from project OpenHFT/Chronicle-Map)  from $Dataset2_{Java}$. This conflict was resolved by the developers by keeping only the line that contains the \texttt{@Override} annotation. MergeGen inferred from the input that one version removed the original content and the other version added a new line to the original content. Thus, it correctly generated the expected resolution. Meanwhile, SBCR generated a candidate containing almost all lines from the bigger version, due to its evaluation function that aims to maximize the similarity with both parents.

\subsubsection{Qualitative Analysis: When SBCR Excels}

Conversely, SBCR demonstrated its strengths in scenarios where MergeGen struggled due to its inherent model and data limitations. Table \ref{tbl:sbcr-sbcr_better_mergegen} summarizes these situations.

\begin{table}[ht]
\centering
\caption{Situations where SBCR performed much better than MergeGen, categorized by dataset.}
\label{tbl:sbcr-sbcr_better_mergegen}
\begin{adjustbox}{max width=\linewidth}
\begin{tabular}{@{}p{5.5cm} rrrrr r@{}}
\toprule
\textbf{Situation} & \textbf{D1} & \textbf{D2\textsubscript{Java}} & \textbf{D2\textsubscript{C\#}} & \textbf{D2\textsubscript{JS}} & \textbf{D2\textsubscript{TS}} & \textbf{Total} \\
\midrule
Empty candidate & 0 & 3 & 3 & 3 & 7 & 16 \\
Non-English content & 1 & 1 & 0 & 4 & 1 & 7 \\
Truncated candidate & 1 & 1 & 4 & 1 & 0 & 7 \\
Other & 2 & 2 & 0 & 2 & 0 & 6 \\
Mismatch with resolution start & 1 & 1 & 0 & 1 & 0 & 3 \\
Unexpected content & 3 & 0 & 0 & 0 & 0 & 3 \\
Unused lines in resolution & 0 & 2 & 0 & 0 & 1 & 3 \\
Malformed chunk & 0 & 0 & 3 & 0 & 0 & 3 \\
Small difference in similarity & 2 & 0 & 0 & 0 & 0 & 2 \\
\midrule
\textbf{Total} & \textbf{10} & \textbf{10} & \textbf{10} & \textbf{10} & \textbf{10} & \textbf{50} \\
\bottomrule
\end{tabular}
\end{adjustbox}
\vspace{0.5em}\\
{\footnotesize \textit{Note:} D1 = $Dataset1$; D2\textsubscript{Java} = $Dataset2_{Java}$; D2\textsubscript{C\#} = $Dataset2_{C\#}$; D2\textsubscript{JS} = $Dataset2_{JavaScript}$; D2\textsubscript{TS} = $Dataset2_{TypeScript}$.}
\end{table}

As shown in the table, the most frequent problem for MergeGen was producing an empty candidate (32\% of cases), particularly when one of the conflicting versions was also empty. Other significant problems included generating truncated candidates due to token limits (14\%) and mishandling non-English content (14\%). In these situations, SBCR's language-agnostic, text-based approach proved to be more robust and capable of generating a complete and more accurate resolution.

\subsubsection{The Impact of Content Balance on Performance}

\begin{figure}[H]
    \centering
    \includegraphics[width=\linewidth]{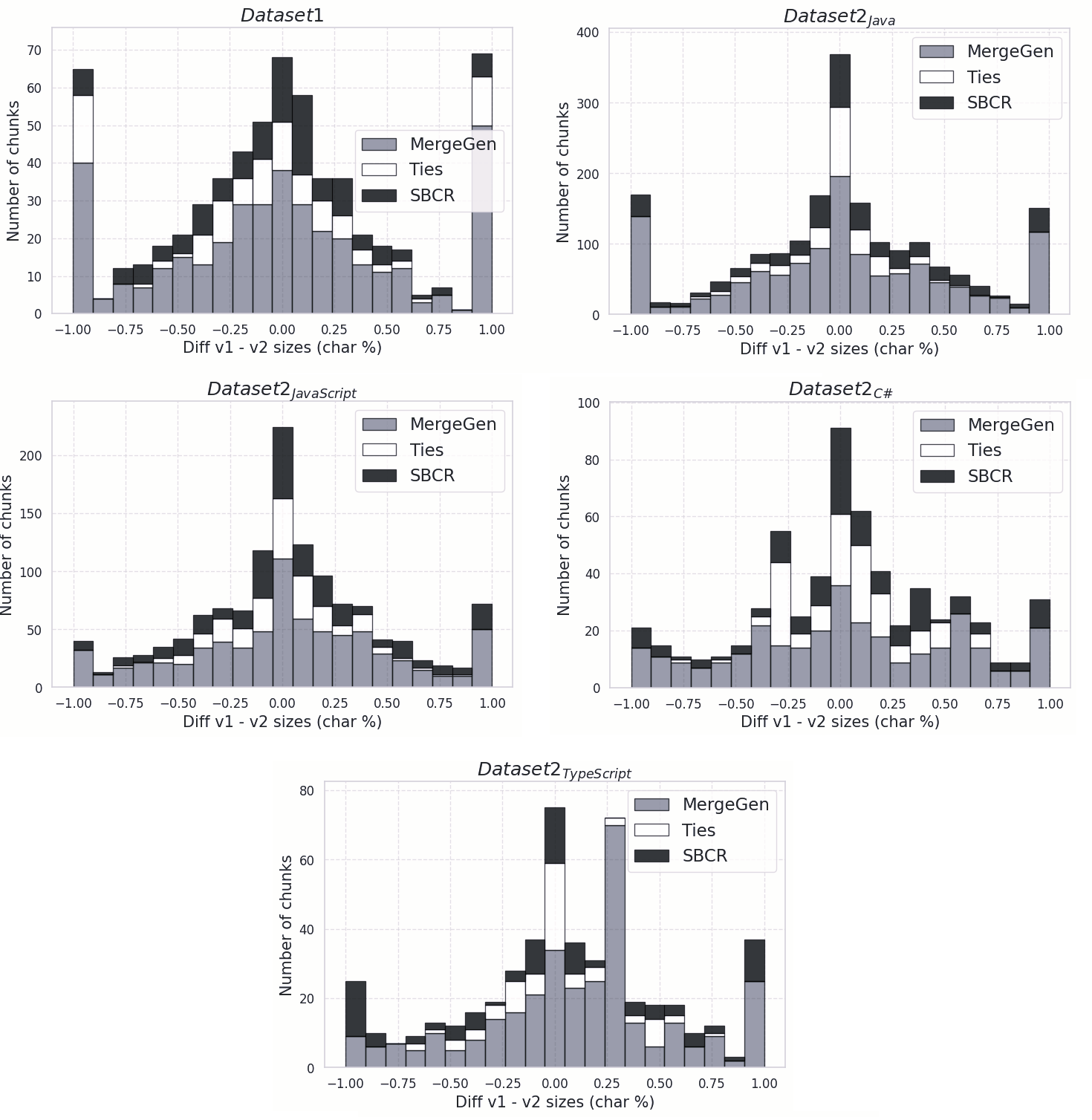}
    \caption[Distribution of the size differences between version 1 ($v_1$) and version 2 ($v_2$) across the analyzed datasets.]{Distribution of the size differences between $v_1$ and $v_2$ when MergeGen is better (gray), when SBCR is better (black), and when they tie (white). Positive values indicate $v_1$ is larger; negative values indicate $v_2$ is larger. The figure illustrates that SBCR performs best on balanced conflicts, while MergeGen excels in imbalanced scenarios.}
    \label{fig:sbcr-rq4_diff_sizes}
\end{figure}

The relationship between performance and the balance of content between conflicting versions is quantified in Figure \ref{fig:sbcr-rq4_diff_sizes}. Dark bars indicate cases where SBCR is better than MergeGen, gray bars indicate cases where MergeGen is better than SBCR, and white bars indicate ties.

\FloatBarrier

The histograms in Figure \ref{fig:sbcr-rq4_diff_sizes} show that SBCR's performance peaks (black bars) when the conflicting versions, $v_1$ and $v_2$, are of similar size (\textit{Diff v1 - v2 sizes} $\approx 0$). In contrast, MergeGen (gray bars) dominates at the extremes of the distribution (\textit{Diff v1 - v2 sizes} $\approx \pm 1$), where one version is significantly larger than the other. This analysis also exposed a case of potential overfitting in MergeGen. In $Dataset2_{TypeScript}$, the distinct gray bar near the 0.25 mark corresponds to a conflict that was repeated multiple times in the training data. This suggests MergeGen may have memorized this specific resolution rather than learning a generalizable strategy, highlighting a key advantage of SBCR's search-based approach, which is not susceptible to this type of overfitting.

\vspace{1em}

\begin{center}
\noindent\fbox{%
    \parbox{0.95\columnwidth}{
        \textbf{Finding 4}: MergeGen excels in scenarios involving unbalanced versions or whitespace content, benefiting from its learned contextual patterns. SBCR, on the other hand, achieves its peak performance with balanced chunks and in situations where MergeGen fails, such as with non-English content or due to model limitations, indicating its strength in generalizing across more diverse situations without overfitting. Each approach has strengths that are context-dependent.
    }
}
\end{center}

\section{Discussion}
\label{sec:discussion}
Our empirical evaluation of MergeGen and SBCR reveals that neither paradigm is universally superior. Instead, they exhibit a series of fundamental trade-offs. In this section, we discuss these trade-offs, the limitations of each approach, and their broader implications for resolving software merge conflicts.

\subsection{Potential Overfitting in MergeGen}
The results achieved by MergeGen raise concerns about overfitting. Our qualitative analysis in RQ4 identified several cases where the quality of the generated resolution seemed remarkably high, yet it was unclear how the approach could deduce such solutions from the conflict data alone. Upon further examination of the training data, we discovered multiple instances of identical or highly similar conflict cases (e.g., the repetitive chunk in the \textit{Azure/autorest.typescript} project), which likely contributed to this high performance. This suggests that the model may have memorized specific examples, which calls into question its generalizability in projects where each merge conflict is unique. Future research should focus on testing MergeGen's capabilities in a live development environment to assess how well it performs on truly unseen conflicts.

\subsection{Handling Conflicts of Different Sizes}
The evaluation highlights varying strengths in handling conflicts of different sizes. As an LLM-based tool, MergeGen requires truncating both the input and output when conflict chunks exceed a configured token limit. This means that for very large conflicts, MergeGen might not provide complete or useful resolutions, limiting its applicability in complex merge situations. In such cases, SBCR presents a viable alternative, as it does not impose size limitations in the same manner. Conversely, for smaller conflicts, MergeGen typically outperforms SBCR due to its contextual understanding, except in the specific failure cases identified in RQ4.

\subsection{Practical Implications of Failure Modes}

Beyond the quantitative similarity scores, the \textit{nature} of the failures from each paradigm has distinct practical implications for developers. Our qualitative analysis in RQ4 provides insights into the different types of risks and cognitive loads associated with each approach.

When the GenAI paradigm (MergeGen) fails, it often does so in a way that is obvious and immediately detectable. Producing an empty or clearly truncated candidate is a loud failure. While this forces the developer to resolve the conflict manually from scratch, thus offering no assistance, it carries a low risk of introducing subtle bugs. The developer is fully aware that the automation has failed and that they must take complete ownership of the resolution process.

In contrast, the SBSE paradigm's (SBCR) typical failure mode can be more subtle and potentially more dangerous. When faced with a highly imbalanced conflict, SBCR may produce a resolution that may appear plausible, yet it may be logically incorrect because its evaluation function forced the balance of lines. A developer under time pressure or with a high degree of trust in the tool might approve such a resolution without deep inspection, potentially introducing a latent bug that is much harder to find later during testing or, in the worst case, in production.

This reveals a critical trade-off in failure modes: the GenAI paradigm, in these specific cases, tends to fail loudly, whereas the SBSE paradigm can fail silently. Understanding this distinction is crucial for designing user interfaces and review processes for automated merging tools.

\subsection{Dependency on the Evaluation Function in SBCR}
A key limitation of SBCR, identified in our qualitative analysis, is its dependency on the evaluation function, particularly its sensitivity to the balance of content between conflicting versions. The current function, which averages the similarity to both versions, performs well on balanced conflicts but struggles when one version is significantly larger. This can lead to suboptimal resolutions in highly imbalanced scenarios. Future work should explore alternative evaluation functions that account for this disparity.

\subsection{Cost, Training, and Practical Considerations}
Finally, a critical distinction between the two paradigms lies in their practical costs. MergeGen, based on a large language model, requires considerable computational resources and financial investment for training and fine-tuning. In our own experiments, resource constraints forced us to truncate inputs, potentially impacting resolution quality. In contrast, SBCR does not require a training phase; its cost is primarily dependent on the configurable search time. This makes SBCR a more cost-effective and accessible solution, particularly for resource-constrained environments. The trade-off is that SBCR's performance is tied to its search time, whereas a fully-resourced and pre-trained MergeGen can generate resolutions almost instantaneously.

\subsection{Towards a Comprehensive Hybrid Approach}

Our findings suggest that the GenAI and SBSE paradigms are not mutually exclusive, but rather complementary. The weaknesses of one approach are often the strengths of the other, which points towards the potential of a hybrid or ensemble tool that intelligently leverages both. Such a system could provide a more robust and effective solution than either paradigm in isolation.

We envision a practical workflow where a ``meta-resolver'' first analyzes the characteristics of an incoming merge conflict. Based on this analysis, it would route the conflict to the most suitable engine:
\begin{itemize}
    \item \textbf{Small to medium-sized, imbalanced conflicts}, or those matching common patterns, would be directed to \textbf{MergeGen}. Its speed and learned contextual understanding make it ideal for these scenarios.
    \item \textbf{Large conflicts} that exceed LLM token limits would be routed to \textbf{SBCR}. Its ability to handle inputs of any size (constrained only by time) makes it a crucial fallback for complex cases where MergeGen would otherwise fail by truncating the input.
    \item \textbf{Conflicts with balanced content} would be prime candidates for \textbf{SBCR}, as our results show it performs optimally in these scenarios where its balancing evaluation function is most effective.
    \item \textbf{Conflicts containing non-English content} or other characteristics known to be outside the typical training distribution of LLMs would be handled by \textbf{SBCR}, leveraging its language-agnostic robustness.
\end{itemize}

This intelligent routing mechanism, potentially implemented as a simple rules-based system or a lightweight classifier, would mitigate the primary weaknesses of each paradigm. It would avoid tasking MergeGen with conflicts it is ill-equipped to handle (thus preventing empty or truncated resolutions) while applying SBCR's more computationally intensive search only when necessary and most effective. This vision of a complementary system aligns with the broader goal of building pragmatic, reliable, and truly automated conflict resolution tools for developers.

\section{Threats to Validity}

This section discusses the potential threats to our study's validity and the measures adopted to mitigate them, covering construct, internal, conclusion, and external validity.

\textbf{Construct Validity} relates to whether our measurements accurately reflect the intended concepts. A primary threat is the assumption that the developer's resolution is the optimal ground truth, which, while common practice, is not guaranteed in terms of correctness or maintainability. Additionally, our main metric, textual similarity, may not fully capture qualitative aspects like code readability. We mitigated this by using a well-established Gestalt-based similarity metric, widely used in prior research, but we acknowledge that similarity alone is not a comprehensive measure of resolution quality.

\textbf{Internal Validity} concerns confounding factors that could affect the results. First, implementation differences could bias time measurements. We measured only MergeGen's generation time, excluding its required pre-processing steps; therefore, its reported time is a lower bound. We have noted this distinction throughout our analysis. Second, LLM-based models are susceptible to overfitting. We mitigated this by using distinct training, validation, and testing partitions for all experiments. Finally, the CodeT5 model underlying MergeGen \citep{wang2021codet5} may have been pre-trained on code from our test datasets, a potential data leakage issue. However, since CodeT5 was not trained specifically on merge scenarios, we believe the impact of this threat is likely limited.

\textbf{Conclusion Validity} relates to the reliability of our findings. To mitigate threats from inappropriate statistical methods or insufficient sample size, we used well-established non-parametric tests (Wilcoxon signed-rank test), the Common Language Effect Size (CLES) metric, and a large dataset comprising thousands of conflicts across multiple programming languages.

\textbf{External Validity} concerns the generalizability of our results. Our findings may not generalize beyond the open-source projects in our datasets (Java, C\#, JavaScript, and TypeScript), as coding practices may differ in proprietary contexts. To mitigate this, we intentionally selected diverse, multi-language datasets that have been used in previous studies, representing a wide variety of conflict scenarios.

\section{Related Work}
\label{sec:related_work}

The challenge of merge conflict resolution has been addressed by a wide range of approaches over the years. To contextualize our comparison of the novel GenAI and SBSE paradigms, this section reviews the main categories of existing work: structured approaches, other optimization techniques, and various learning-based methods.

\subsection{Structured and Semi-Structured Merge Approaches}

Early attempts to improve upon purely textual merging focused on leveraging the code's structure through semi-structured and structured approaches. These methods aim to improve conflict resolution effectiveness by using language-specific information, overcoming some limitations of unstructured, line-based merging.

\citet{apelSemistructuredMergeRethinking2011k} introduce the concept of semi-structured merge as a middle ground between unstructured and structured systems. To balance the lack of semantic awareness in unstructured tools (like Git's default) and the implementation overhead of fully structured tools, they propose FSTMERGE, which uses annotated grammars to introduce structural awareness. Their evaluation shows a significant reduction in conflicts, though handling some cases like renaming remains challenging. Building on this, \citet{apelStructuredMergeAutoTuning2012a} present JDime, a structured merge approach that dynamically switches between structured and unstructured merging to balance precision and performance.

Later work by \citet{cavalcantiEvaluatingImprovingSemistructured2017o} and \citet{cavalcanti2024semistructured} further advanced these ideas, demonstrating that leveraging structural information, even partially, significantly reduces false positive conflicts, making them easier for developers to analyze.

Compared to these approaches, which are tightly coupled to language-specific parsers, both paradigms in our study offer different forms of flexibility. The SBSE paradigm, represented by SBCR, is inherently language-agnostic due to its textual nature. The GenAI paradigm, represented by MergeGen, while requiring language-specific training data, can be adapted to any language where such data is available, offering a broader potential application than tools with hardcoded grammars.

\subsection{Combination and Optimization Approaches}
A different line of work has focused on optimization techniques, treating conflict resolution as a problem of finding the best combination of changes. These approaches vary in their reliance on historical data and language-specific features.

For instance, \citet{zhu2018conflict} proposed AutoMerge, which uses version space algebra (VSA) and ASTs to efficiently represent and rank a large set of candidate resolutions. \citet{xing2019automatic} explored Automated Program Repair (APR) techniques, using genetic programming to generate fixes for behavioral conflicts. More aligned with learning, \citet{gonzalez2022almost} presented ``Almost Rerere'', which uses genetic programming to generate resolution rules from conflicts resolved in the past. While powerful, these approaches often depend on language specificity (AutoMerge) or the availability of high-quality historical examples (Almost Rerere).

The SBSE paradigm for conflict resolution, as represented by SBCR in this paper, is built upon two key foundational findings from our prior work. First, \citet{camposjunior2024composition} established heuristics for efficiently navigating the solution space, finding that 98.6\% of combination-based resolutions maintain the partial order of the original lines. This insight provides a principled way to drastically reduce the search space. Second, \citet{camposjunior2025sbcr} addressed the challenge of evaluating candidates by proposing a lightweight, similarity-based evaluation function. The study found that developer resolutions are, on average, 70\% similar to both parent versions and, crucially, that a candidate's similarity to its parents is strongly correlated ($\rho=0.791$) with its similarity to the final resolution \citep{camposjunior2025sbcr}. This provides a feasible proxy for solution quality, avoiding expensive compilation and testing. Taken together, these findings provide the two necessary components for an SBSE approach: a constrained search space and a feasible evaluation function. While sharing the goal of optimization with SBCR, many of the other approaches in this category, such as Almost Rerere, depend on historical data or language-specific features like ASTs. The SBSE paradigm, as implemented in SBCR, distinguishes itself by using these empirically-derived, data-agnostic heuristics, offering a more generalizable optimization strategy for novel scenarios.

\subsection{Learning-Based Approaches}
More recently, the advent of large-scale data and deep learning has given rise to a new category of learning-based approaches, which forms one of the core paradigms of our study. These methods leverage neural networks and large language models (LLMs) to improve accuracy.

Early models like DeepMerge \citep{dinella2022deepmerge} used edit-aware embeddings to construct resolutions, while MergeBERT \citep{svyatkovskiy2022program} framed the task as a token-level classification problem. A significant leap came with generative models. Gmerge \citep{zhang2022using} applied large pre-trained models like GPT-3 to the task, while MergeGen \citep{dong2023merge} moved away from classification entirely. By treating conflict resolution as a pure generation task, MergeGen can produce novel code and combinations beyond predefined patterns, establishing it as the state-of-the-art.

These learning-based techniques represent the paradigm exemplified by MergeGen in our study. They leverage historical data and complex models to understand semantic aspects of conflicts. This stands in direct contrast to the SBSE paradigm (SBCR), which applies a heuristic-based search without requiring historical data or deep learning models. While tools like MergeGen excel in conflicts that require generative capacities, SBCR provides a more computationally feasible and data-independent solution, particularly appealing where training data is unavailable or resources are limited.

\subsection{Comparison and Evaluation of Conflict Resolution Approaches}
Evaluating the effectiveness of these diverse approaches is a significant challenge in itself. To address this, researchers have developed dedicated benchmarks. \citet{shen2024conflictbench} introduced ConflictBench, a standardized dataset of Java merging scenarios to systematically compare tools on metrics like precision and resolution desirability. In a complementary effort, \citet{schesch2024evaluation} conducted an in-depth evaluation of multiple tools, emphasizing merge correctness through test suite execution. Their work highlights that tool performance is highly scenario-dependent.

In contrast to these broad evaluations of multiple tools, such as ConflictBench \citep{shen2024conflictbench} and the work by \citet{schesch2024evaluation}, our study provides a focused, in-depth comparison of two distinct and novel paradigms for merge conflict resolution: search-based optimization (SBCR) and Generative AI (MergeGen). Our review of the literature reveals a clear evolution in this field, from early structured approaches to more recent learning-based techniques. However, it also highlights persistent gaps, such as limitations in language-specificity and a heavy dependency on historical data. To our knowledge, no prior work has conducted a direct empirical comparison of the emerging GenAI and SBSE paradigms. Our study addresses this critical gap by analyzing their trade-offs in adaptability, performance, and practical failure modes, aiming to deepen the understanding of how these modern approaches may complement each other.

\section{Conclusion}
\label{sec:conclusion}

This paper presented the first in-depth empirical comparison between two competing and novel paradigms for merge conflict resolution: Generative AI, represented by MergeGen, and Search-Based Software Engineering (SBSE), represented by SBCR. We evaluated both paradigms across multiple datasets consisting of Java, C\#, JavaScript, and TypeScript codebases. The experiments included a comprehensive parameter tuning process for SBCR and a systematic performance evaluation of both approaches in terms of resolution similarity, execution time, and generalizability. We also performed a qualitative analysis to understand the specific scenarios where each paradigm excels.

Our findings reveal that the SBSE and GenAI paradigms each have distinct strengths and limitations. The SBSE approach (SBCR) achieved its best performance for conflicts with balanced content between versions. In contrast, the GenAI approach (MergeGen) demonstrated superior results in cases where a strong imbalance existed, potentially benefiting from learned behavior in its training data. We also identified key weaknesses: SBCR's performance is sensitive to its evaluation function in highly unbalanced scenarios, while MergeGen showed clear signs of overfitting.

A key conclusion from our study is that, for practical conflict scenarios, the ideal resolution strategy involves a hybrid approach that leverages the complementary strengths of both paradigms. MergeGen performed effectively for smaller and more predictable conflicts, while SBCR showed advantages for larger, more complex scenarios, or those with characteristics (like non-English content) that fall outside the LLM's typical training data. Thus, a combined approach that routes conflicts based on their characteristics could provide developers with a more robust and effective solution.

The implications of these results suggest that relying on a single paradigm for conflict resolution may not be sufficient in real-world software development. A data-dependent approach like MergeGen may struggle with novel conflicts, while a generalized search-based solution like SBCR can complement these limitations. Our findings highlight the need for more adaptive and hybrid solutions that intelligently select the appropriate technique, thus reducing manual intervention and increasing developer productivity.

For future work, we plan to explore alternative evaluation functions for SBCR to improve its performance in unbalanced scenarios. We also intend to conduct an experimental evaluation of the proposed hybrid system in a real-world development setting, involving developers to gain insights into the usability, practicality, and perceived effectiveness of each paradigm in daily development tasks. These studies are crucial steps towards moving beyond the comparison of individual paradigms and into the engineering of a new generation of intelligent, context-aware merging tools. By synergistically combining the generative power of LLMs with the robust, heuristic-driven search of SBSE, such tools have the potential to significantly reduce developer burden and fundamentally improve collaborative software development workflows.

\section*{Declarations}

\subsection*{Funding}
We thank the following entities for partially funding this work: INES.IA (National Institute of Science and Technology for Software Engineering Based on and for Artificial Intelligence) www.ines.org.br; Conselho Nacional de Desenvolvimento Cient\'{i}fico e Tecnol\'{o}gico (CNPq) for the grants 309300/2023-1, 408817/2024-0, and 155576/2025-9; and Funda\c{c}\~{a}o Carlos Chagas Filho de Amparo \`{a} Pesquisa do Estado do Rio de Janeiro (FAPERJ) for the grant E-26/204.145/2024.

\subsection*{Conflicts of interest/Competing interests}
The authors have no relevant financial or non-financial interests to disclose.

\section*{Data Availability}

In the spirit of open science and to ensure the full reproducibility of our study, we have made all data, scripts, and results publicly available in a comprehensive replication package.

The main repository, containing the source code for both the SBCR and MergeGen approaches, along with all scripts and detailed instructions to replicate the experiments and analyses, is available on GitHub at:
\begin{center}
    \url{https://github.com/gems-uff/sbcr_study}
\end{center}

To further facilitate replication and inspection of our results, we also provide two archival datasets on FigShare:

\begin{itemize}
    \item \textbf{Pre-processed Datasets:} For researchers who wish to bypass the initial data pre-processing steps, the final, clean datasets used directly in our experiments are available at: \url{https://figshare.com/s/d196f4ccb3ef34d2e770}.

    \item \textbf{Full Experimental Results:} A complete archive of our experimental run, including intermediate files, execution logs, all candidates generated for each conflict, and the models generated by MergeGen, can be found at: \url{https://figshare.com/s/b3cdd351d077a9b08121}. This allows for a deep inspection of all generated artifacts.
\end{itemize}

\subsection*{Authors' Contributions}
All authors contributed to the study conception and design. Material preparation and data collection was performed by Heleno. Data analysis was performed by Heleno and Leonardo. The first draft of the manuscript was written by Heleno and all authors commented on previous versions of the manuscript. All authors read and approved the final manuscript.

\bibliographystyle{plainnat}
\bibliography{references}

\end{document}